\documentclass[graybox]{svmult}

\usepackage{type1cm}        
\usepackage{graphicx}        

\usepackage[bottom]{footmisc}

\usepackage{newtxtext}       %
\usepackage{newtxmath}       

\usepackage{bm}

\usepackage{url}
\usepackage{amsmath}
\usepackage{booktabs}

\begin{document}

\title*{Machine Learning Applications in Gravitational Wave Astronomy}
\author{Nikolaos Stergioulas}
\institute{Nikolaos Stergioulas \at Department of Physics, Aristotle University of Thessaloniki, 54124 Thessaloniki, Greece \email{niksterg@auth.gr}}
%
%
\maketitle

\abstract{ 
Gravitational wave astronomy has emerged as a new branch of observational astronomy, since the first detection of gravitational waves in 2015. The current number of $O(100)$ detections is expected to grow by several orders of magnitude over the next two decades. As a result, current computationally expensive detection algorithms will become impractical. A solution to this problem, which has been explored in the last years, is the application of machine-learning techniques to accelerate the detection and parameter estimation of gravitational wave sources. In this chapter, several different applications are summarized, including the application of artificial neural networks and autoenconders in accelerating the computation of surrogate models, deep residual networks in achieving rapid detections with high sensitivity, as well as artificial neural networks for accelerating the construction of neutron star models in an alternative theory of gravity. }

\section{Introduction}
\label{sec:1}

\noindent  Since 2015, when the first gravitational waves (GWs) from a binary black hole (BBH) system were detected \cite{Abbott:2016blz}, GW detections have become increasingly common, moving closer to the point of being a regular occurrence. After the third observing run (O3), the most recent catalog (GWTC-3, \cite{GWTC3_LVK_2021}) from the Advanced LIGO \cite{TheLIGOScientific:2014jea}, Advanced Virgo \cite{acernese2014advanced} KAGRA \cite{akutsu2019kagra, KAGRA_2021} collaboration contained 90 GW events, almost all of which were BBH mergers. The 4th observing run (O4) is currently underway and a larger number of BBH detections are expected \cite{abbott2020prospects}.
The addition of a fifth interferometer, LIGO-India \cite{LIGO_India_2022}, is expected to significantly enhance both the sensitivity and the sky localization of the network. Moreover, third-generation ground-based detectors such as the Einstein Telescope \cite{Punturo_etal_2010_ET, Maggiore_etal_2020_ET} and Cosmic Explorer \cite{Reitze_etal_2019_CE, Evans_etal_2021_CE} are currently being developed and are anticipated to greatly expand our understanding of the astrophysical processes in the Universe \cite{NextGen_detectors_2017, GWIC_3G_reports_intro_2021, GWIC_3G_reports_science_2021}.

The advances in GW astronomy described above were made possible by collaborative efforts in multiple areas. Accurate descriptions of the entire coalescence, including the full inspiral, merger, and ringdown, can be obtained in different ways, with \texttt{IMRPhenomXPHM} \cite{pratten2021IMRPhenomXPHM} and \texttt{SEOBNRv5PHM} \cite{2023arXiv230318046R} being two examples of waveform models. Recent implementations of these models take into account the spin-induced precession of the binary orbit and contributions from both the dominant and subdominant multipole moments of the emitted gravitational radiation. However, the increased complexity of the waveforms increases their computational cost. 

Astronomical observations have enabled a number of attempts to determine the Equation of State (EoS) of Neutron Stars (NSs). These include the NICER mass and radius measurements \cite{Riley_2019,Miller_2019,Miller_2021}, the measurement of tidal deformability through gravitational waves \cite{Van_Oeveren_2017,Hinderer:2007mb,Chatziioannou2020,Dietrich2021}, as well as joint constraints, e.g., \cite{Biswas_2022,Tim_Dietrich,Landry_2020PhRvD.101l3007L,Raaijmakers:2021uju}. In particular, the detection of the binary NS merger GW170817 \cite{LIGOScientific:2017vwq,LIGOScientific:2018hze} has prompted further research in this area.

In recent years, there has been an increase in the utilization of machine learning approaches for the analysis of gravitational wave data (see \cite{cuoco2020review,app13179886,2023arXiv231115585Z} for reviews). This chapter provides a summary of different machine learning applications to gravitational-wave astronomy presented in \cite{2022arXiv220308434F,2022Neurc.491...67N,2023PhRvD.108b4022N,2023arXiv230903991L}.

\section{ANN-Accelerated Surrogate Models}
\label{sec:surrogate}

Surrogate modeling has been provided to reduce the considerable computational cost of evaluating waveform models \cite{field2014fast, Tiglio_Villanueva_2021_review_arXiv210111608T}, which can significantly speed up EOB waveforms (e.g. \cite{field2014fast, Purrer_2016, Lackey_etal_2017, Lackey_etal_2019, Yun_etal_2021}) while still providing high accuracy within its valid parameter range.
The SEOBNRv4 model has a three-dimensional parameter space $\boldsymbol{\lambda}$; the mass ratio $q$ between the two black holes and their spins $\chi_1$ and $\chi_2$, assuming that they are aligned with the orbital angular momentum. A surrogate model for this waveform family was presented in \cite{khan2021gravitational}. Several machine learning techniques can be used to interpolate or fit the projection coefficients of a reduced basis representation of time-domain waveforms, and the most suitable method depends on the desired accuracy and dimensionality. For low-dimensional parameter spaces, interpolation is a viable option. However, as the dimensionality increases, interpolation becomes difficult due to the large number of data points usually needed. Artificial Neural Networks (ANNs) are proposed as a solution to estimate these coefficients since this approach allows for efficient execution on either a CPU or GPU.

In \cite{2022arXiv220308434F}, it was observed that the residual errors after training an ANN to evaluate the coefficients of the surrogate model for the SEOBNRv4 model had a pattern with respect to the input parameters. It was then demonstrated that a second neural network could be trained to model these errors, leading to an improved method, in which the maximum mismatch between SEOBNRv4 waveforms and waveforms generated by the new surrogate model was more than one order of magnitude smaller than the baseline method. Here, we will provide a summary of the steps taken to create the surrogate model and the residual ANN network to accelerate the evaluation of its coefficients, as described in \cite{2022arXiv220308434F}.

\subsection{Constructing a surrogate model}
We express the complex gravitational wave strain as $h(t;\boldsymbol{\lambda}) = h_+(t;\boldsymbol{\lambda}) - ih_{\times}(t;\boldsymbol{\lambda})$, where $h_+$ and $h_{\times}$ are the two independent polarizations \cite{maggiore}, $t$ is the time, and $\boldsymbol{\lambda}$ is a vector of intrinsic parameters. The SEOBNRv4 model \cite{bohe2017improved} has a three-dimensional parameter space, with each waveform characterized by the mass ratio $q$ (the ratio of the masses of the two black holes) and the dimensionless spins $\chi_1, \chi_2$ of the two black holes. Surrogate modeling is a process of approximating given signals using a reduced model, denoted $h_{s}(t;\boldsymbol{\lambda})$, such that the approximation given by the surrogate model, $h_{s}(t;\boldsymbol{\lambda})$, accurately reconstructs the actual waveform $h(t;\boldsymbol{\lambda})$ within a preset threshold of error. When considering only the dominant, quadrupole ($l=m=2$) mode \cite{maggiore}, the target becomes $h_{s}(t;\boldsymbol{\lambda}) \approx h_{2,2}(t;\boldsymbol{\lambda})$ where $l,m$ are the spherical harmonics. To begin the surrogate modeling process, a training set of $N$ waveforms $\{ h_i(t;\boldsymbol{\lambda}_i) \}_{i=1}^{N}$ is created, where ${\boldsymbol{\lambda}}_i = (q, \chi_1, \chi_2)_i$. The mass ratio is limited to a predetermined interval, such as $1\leq q \leq 8$, within which the surrogate model is designed to be accurate. The two spins can have values in the range $-0.99\leq \chi_{1,2} \leq 0.99$.

A \emph{Reduced Order Method} (ROM) basis is constructed from a training set using a greedy algorithm \cite{field2014fast}. This is an iterative process that selects $n < N$ waveforms (and their corresponding ${\{\boldsymbol\lambda}_j\}_{j=1}^{n}$ values, the greedy points) that, after orthonormalization, form the reduced basis $\{e_j\}_{j=1}^{n}$. Each $\boldsymbol{\lambda}_i$ waveform in the training set is then expressed as a linear combination \begin{equation} h(t;\boldsymbol{\lambda}_i) \approx \sum_{j=1}^{n} c_j(\boldsymbol{\lambda}_i) e_j(t), \end{equation} within a given error tolerance, where $\{c_j(\boldsymbol{\lambda}_i)\}_{j=1}^{n} = \left\langle h(t ; \boldsymbol{\lambda}_i), e_{j}(t)\right\rangle$ are the orthogonal projection coefficients.

Next, a new \emph{Empirical Interpolation Method} (EIM) basis $B_k(t)$ is obtained such that a waveform $h(t;\boldsymbol{\lambda}_j)$ can be expressed as a linear combination of the basis, i.e. $h\left(t ; \boldsymbol{\lambda}_j\right)=\sum_{k=1}^{n} \alpha_k(\boldsymbol{\lambda}_j) B_{k}(t)$. The coefficients $\alpha_k(\boldsymbol{\lambda}_j)$ are equal to the waveform at particular times, $\{ T_k \}_{k=1}^{n}$, known as the empirical time nodes, i.e. $\alpha_k(\boldsymbol{\lambda}_j) =h\left(T_{k} ; \boldsymbol{\lambda}_j \right)$. For any other waveform $h\left(t ; \boldsymbol{\lambda}_i\right)$ in the training set, the coefficients of the EIM representation are $\alpha_k(\boldsymbol{\lambda}_i) =h\left(T_{k} ; \boldsymbol{\lambda}_i \right)$. This does not require the basis $B_k(t)$, so the coefficients can be computed much faster than the projection coefficients in the ROM basis (which require the projection of the whole waveform).

In the end, a surrogate model is created by interpolating the coefficient matrix $\alpha_k(\boldsymbol{\lambda}_i)$ of the training set to find the coefficients $\hat \alpha_k(\boldsymbol{\lambda})$ for any $\boldsymbol{\lambda}$, such that 
\begin{equation} h\left(t ; \boldsymbol{\lambda}\right)\approx \sum_{k=1}^{m} \hat \alpha_{k}(\boldsymbol{\lambda}) B_{k}(t). \label{eq:EIM} \end{equation} 
The complexity of this process increases with the number of parameters in $\boldsymbol{\lambda}$. {\it Neural networks can be used to speed up this part of the process, as demonstrated in} \cite{khan2021gravitational}.

In practice, the complex waveform can be expressed in terms of its {\it amplitude} $A$ and {\it phase} $\phi$, defined through 
\begin{equation}
h_{+}(t ; \boldsymbol{\lambda})-h_{\times}(t ; \boldsymbol{\lambda})=A(t ; \boldsymbol{\lambda}) e^{-i \phi(t ; \boldsymbol{\lambda})},
\end{equation}
which leads to a  more compact EIM basis. To construct the ROM and EIM bases, a training set of $N=2 \times 10^5$ waveforms was randomly sampled in the parameter space of $1\leq q\leq 8, -0.99\leq \chi_{1,2}\leq 0.99$. The waveforms were aligned in amplitude and initial phase, the phase was unwrapped, and the time series was truncated to a common starting time of $-20000M$, with a total mass of $M=60M_\odot$. This ensured that all waveforms began with a minimum frequency no larger than 15 Hz, and $100M$ of post-peak ringdown data was kept. The ROM and EIM bases were created using {\tt RomPy} \cite{field2014fast,rompy}. To evaluate the accuracy of the reconstructed waveforms (after the training is completed), a {\it validation set} of $3\times 10^4$ SEOBNRv4 waveforms (not included in the training set) was used.

For two waveforms with parameters $\boldsymbol{\lambda}_1$ and $\boldsymbol{\lambda}_2$, the inner product can be defined \cite{flanagan}
\begin{equation}
    \langle h(\cdot; \boldsymbol{\lambda}_1), h(\cdot; \boldsymbol{\lambda}_2) \rangle = 4\Re\int_{f_{min}}^{f_{max}} \frac{\tilde h(f;\boldsymbol{\lambda}_1) \tilde h^{*}(f;\boldsymbol{\lambda}_2)}{S_{n}(f)} df,
\end{equation}
where $\tilde h(f;\boldsymbol{\lambda})$ is the Fourier transform of $h(t;\boldsymbol{\lambda})$, $S_{n}(f)$ denotes the noise power spectral density (PSD) of the GW detector and the star notation stands for the complex conjugate. The inner product can be employed to normalize the Fourier transform of a waveform in the following manner:
\begin{equation}
    \hat{h}(f; \boldsymbol{\lambda}) = \frac{{\tilde h}(f; \boldsymbol{\lambda})}{\langle h(\cdot; \boldsymbol{\lambda}), h(\cdot; \boldsymbol{\lambda}) \rangle},
\end{equation}
Then, the {\it overlap} between two waveforms is defined as the inner 
product between normalised waveforms $\hat{h}(\cdot; \boldsymbol{\lambda}_1)$, $\hat{h}(\cdot; \boldsymbol{\lambda}_2)$, maximised over a relative time ($t_0$) and phase ($\phi_0$) shift between
the two waveforms:
\begin{equation}
    {\cal O}(\hat{h}(\cdot; \boldsymbol{\lambda}_1), \hat{h}(\cdot; \boldsymbol{\lambda}_2)) = \max_{t_0, \phi_0} \langle h(\cdot; \boldsymbol{\lambda}_1), h(\cdot; \boldsymbol{\lambda}_2) \rangle,
\end{equation}
and, finally the {\it mismatch} is given by
\begin{equation}
    \label{eq:mismatch}
    \mathcal{M}(\hat{h}(\cdot; \boldsymbol{\lambda}_1), \hat{h}(\cdot; \boldsymbol{\lambda}_2)) = 1 - {\cal O}(\hat{h}(\cdot; \boldsymbol{\lambda}_1), \hat{h}(\cdot; \boldsymbol{\lambda}_2)) .
\end{equation}
The performance of the surrogate model can  be evaluated by comparing the waveforms generated by the SEOBNRv4 model with the predictions of the surrogate, using the mismatch defined above.

\begin{figure}[ht!]
	\includegraphics[width=12cm]{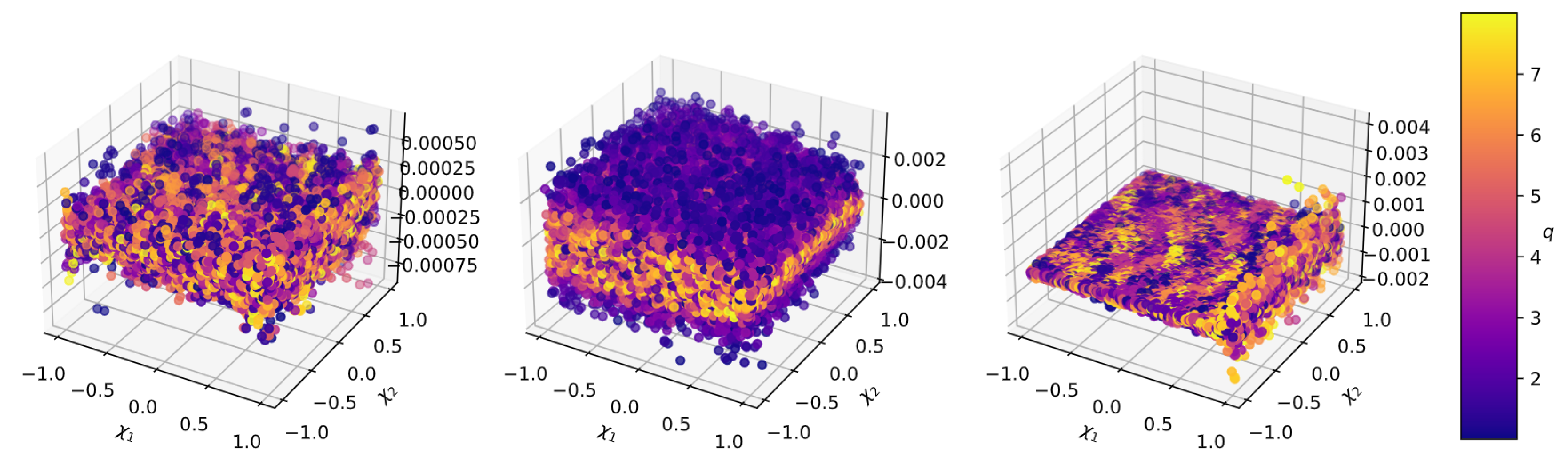}
	\centering
	\caption{The residual error for three chosen EIM coefficients for the amplitude is dependent on the input parameters $\boldsymbol{\lambda}=\{\chi_1, \chi_2, q\}$ (the dependence on $q$ is illustrated with a colormap). The example in the left panel shows an unstructured distribution of residuals. In contrast, the example in the center panel reveals a strong dependence on the mass ratio $q$, while the example on the right displays a large residual error at the highest value of $\chi_1$. Figure from \cite{2022arXiv220308434F}.}
	\label{amp errors}
\end{figure}

\subsection{Accelerating the surrogate model using ANNs}

To construct the surrogate model, an ANN was employed to interpolate the coefficients $\alpha_k({\boldsymbol{\lambda}_i})$ of the training set to find the coefficients $\hat \alpha_k(\boldsymbol{\lambda})$ for an arbitrary $\boldsymbol{\lambda}$. The improved model was compared with a baseline model that followed the architecture of \cite{khan2021gravitational}. The ANN had four hidden layers with 320 neurons in each. The batch size was $10^{3}$ and the training lasted for $10^{3}$ epochs. The Adam optimizer \cite{adam} with a learning rate of $10^{-3}$ and the ReLU activation function \cite{relu} were used for the amplitude network. For the phase network, the Adamax \cite{adam} optimizer with a learning rate of $10^{-2}$ and the softplus activation function \cite{softplus} were employed. Preprocessing involved using $\log(q)$ as input instead of $q$, which was then scaled using the {\tt StandardScaler} from {\tt Scikit-Learn} \cite{sklearn}. At the output, the coefficients were used raw for the amplitude network and were scaled using {\tt Scikit-Learn}'s {\tt MinMaxScaler} for the phase network. 

The ANN prediction of the EIM coefficients of the training set waveforms will be referred to as $\hat{\boldsymbol{y}}_i \equiv \{\hat{\alpha}_{k}(\boldsymbol{\lambda}_i)\}_{k=1}^{n}$. During training, the standard mean square error
\begin{equation}
    MSE = \frac{1}{N} \sum_{i=1}^{N} \| \hat{\boldsymbol{y}}_i - \boldsymbol{y}_i \|_2^2
\end{equation}
was measured and minimized, where the $\| \cdot \|_2$ notation represents the Euclidean norm of a vector. The MSEs were in the range $\sim 10^{-8}-10^{-7}.$

\begin{figure}[ht!]
	\includegraphics[width=7cm]{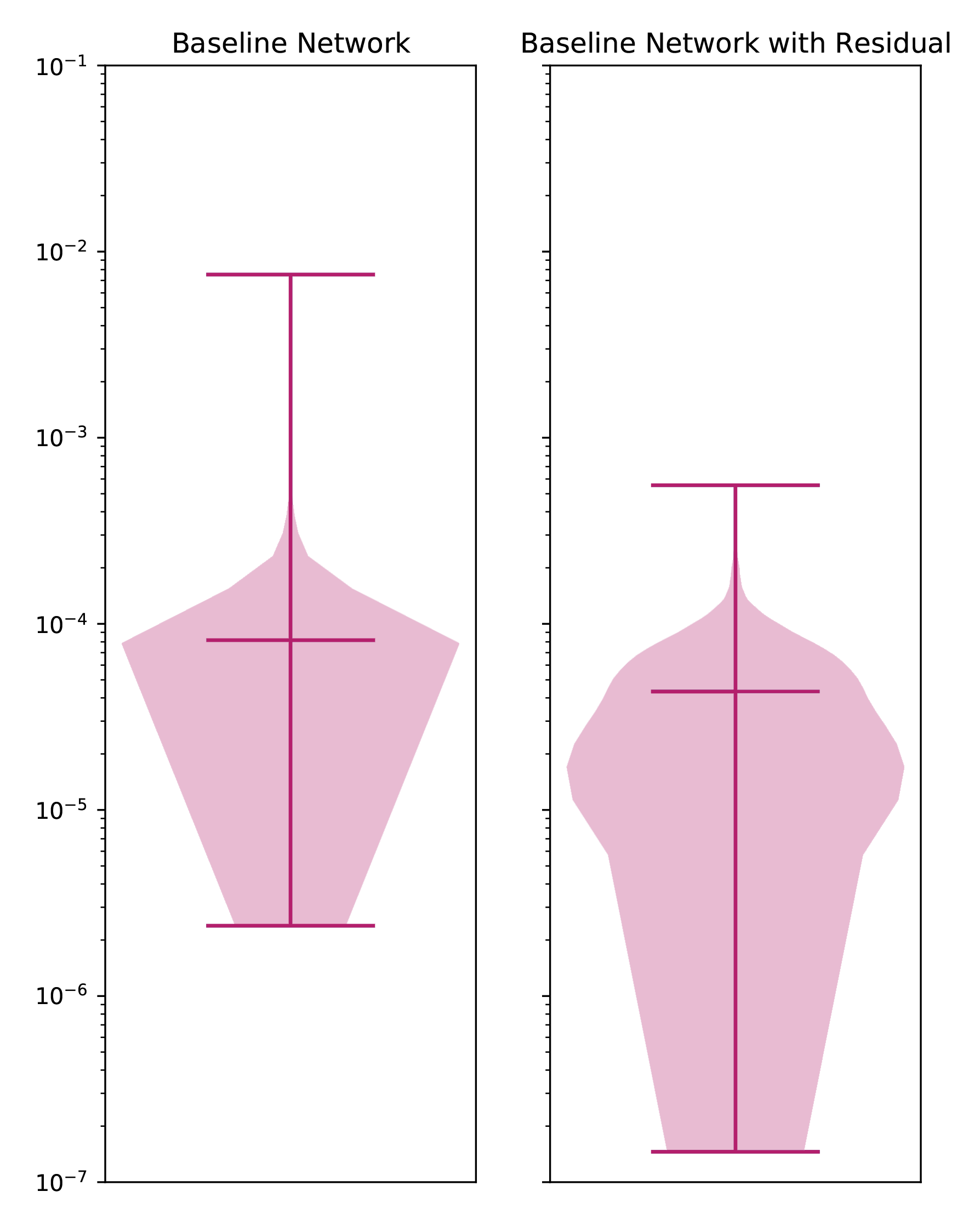}
	\centering
	\caption{A comparison of the mismatches (for the validation set) between the baseline network and the case when a second network that models the residual error is added is shown in the violin plots. The median is marked by the middle horizontal line, while the minimum and maximum values are indicated by the extent of the lines. The envelope of each panel is proportional to the density of points, and it is clear that a significant reduction of the mismatch is achieved when the network for the residual error is added.  Figure from \cite{2022arXiv220308434F}.}
	\label{res-violin}
\end{figure}

A second ANN was created to predict the residual errors after establishing the baseline ANN surrogate model. This was done due to the presence of structure in the residuals for some EIM coefficients, as seen in Fig. \ref{amp errors}. The final predictions are the sum of the outputs of the two models.
For all $\{\boldsymbol{\lambda}_i\}_{i=1}^{N}$ in the training set, one can obtain the corresponding predictions  \{$\hat{\boldsymbol{y}}(\boldsymbol{\lambda}_i)\}_{i=1}^{N}$ and calculate the {\it residual} 
\begin{equation}
    \label{eq:residual_errors}
    \boldsymbol{e}_i \equiv \boldsymbol{y}(\boldsymbol{\lambda}_i) - \hat {\boldsymbol{y}}(\boldsymbol{\lambda}_i),
\end{equation}
where, as already defined, $\boldsymbol{y}$ is the ground truth.  The second network was created with the {\it same input and architecture as the first network}, but this time it was trained on the residuals $\boldsymbol{e}_i$ (which were first scaled using the ``MinMaxScaler'' from \texttt{scikit-learn} \cite{sklearn}) to make predictions for the residual $\hat {\boldsymbol{e}}(\boldsymbol{\lambda}) $ at any $\boldsymbol{\lambda}$.
When the prediction $\hat {\boldsymbol{e}}$ for the residual is added to the prediction $\hat{\boldsymbol{y}}$ of the first network, an improved prediction is obtained.
\begin{equation}
    \tilde{\boldsymbol{y}} \equiv \hat{ \boldsymbol{y}} +  \hat{\boldsymbol{e}}.
\end{equation}

Fig. \ref{res-violin} illustrates the difference in mismatches (for the validation set) between the baseline network and the case where a second network is added that models the residual error, as a violin plot. The median is marked by the middle horizontal line, whereas the minimum and maximum values are shown by the extent of the lines. The envelope of each panel is proportional to the density of points. The results in Fig. \ref{res-violin} demonstrate that adding a second network to learn the residual errors is beneficial for constructing surrogate models for gravitational waves from BBH inspiral. This strategy is likely to be advantageous for other types of GW template banks, such as binary neutron star inspiral waveforms.

\section{Efficient Surrogate Models using Autoencoders}

\begin{figure}
    \centering
    \includegraphics[width=0.9\linewidth]{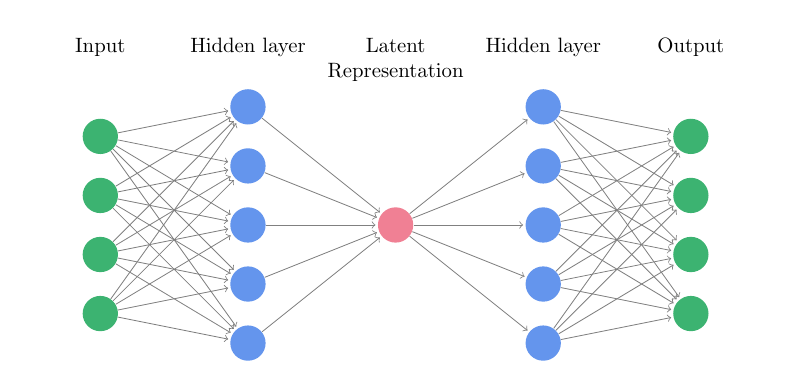}
    \caption{Single hidden layer architecture of a fully connected Autoencoder. Figure from \cite{2022Neurc.491...67N}.}
    \label{fig:ae_architecture}
\end{figure}

Autoencoders (AEs) are a type of unsupervised neural network that is trained to reproduce its input by first transforming it into a lower-dimensional representation~\cite{vincent2008extracting}. Generally, an autoencoder consists of an encoding component that maps the input to a compressed representation and a decoding component that reconstructs the input. Encoding and decoding functions can have symmetrical or asymmetrical architectures and usually comprise multiple layers of fully connected layers, convolutional layers, or recurrent modules. AEs have been studied for a variety of tasks, such as clustering~\cite{xie2016unsupervised,nousi2018self}, classification~\cite{nousi2017deep,nousi2017discriminatively}, and image retrieval~\cite{wu2013online,carreira2015hashing}, due to their ability to extract semantically meaningful representations without labels. A typical AE architecture is shown in Figure~\ref{fig:ae_architecture}, with the input and output layers having the same number of neurons. 

\begin{figure}
    \centering
    \includegraphics[width=0.7\textwidth]{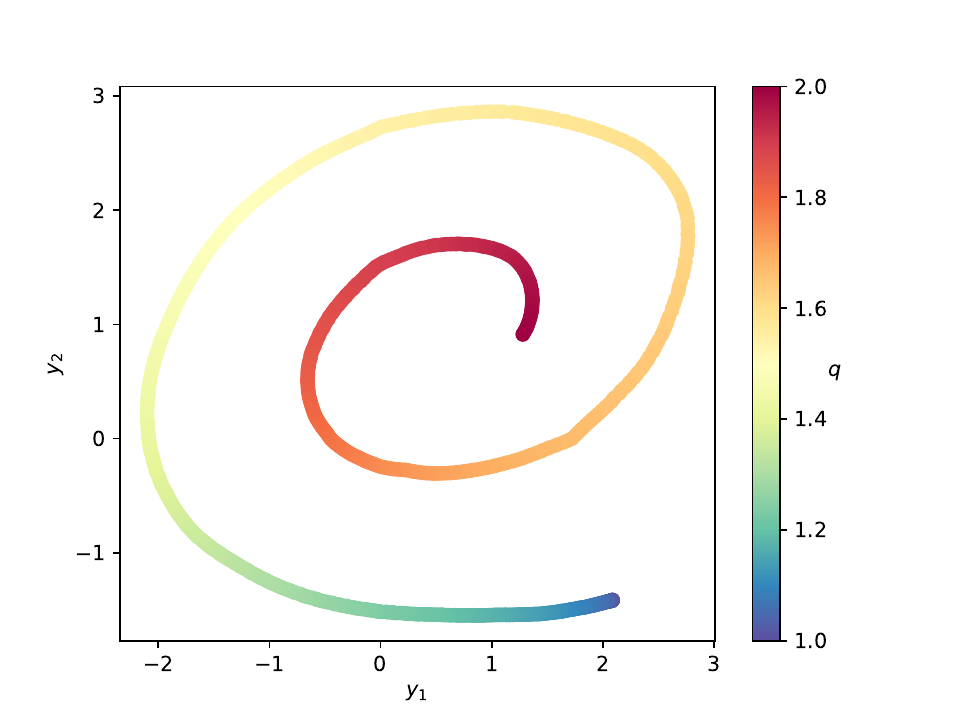}
    \caption{Hidden representation uncovered by the AE for the empirical interpolation coefficients of a surrogate model of EOBNRv2 waveforms that is valid for $1\leq q \leq 2$ (the color bar describes different values of the mass ratio $q$). When the values at the two neurons of the hidden representation are plotted against each other, a spiral structure emerges, along which the mass ratio appears to vary linearly with angle. Figure from \cite{2022Neurc.491...67N}.}
    \label{fig:ae_hidden_rep}
\end{figure}

In \cite{2022Neurc.491...67N} a dataset comprising pairs of mass ratio $q_i$ and corresponding EIM coefficient $\mathbf{a}_i$  ($i=1,...,N$), created with the EOBNRv2 nonspinning waveform model \cite{2011PhRvD..84l4052P}, was used to train an AE, with only the coefficients as input. This unsupervised process revealed a hidden relationship between each mass ratio $q_i$ and the corresponding coefficients, as the mass ratios were unknown to the AE. Specifically, when choosing a two-dimensional intermediate representation, a spiral pattern emerged when visualizing this representation as a function of the mass ratio $q$, see Figure~\ref{fig:ae_hidden_rep}. Below, we summarize the main steps presented in \cite{2022Neurc.491...67N} to add a learnable spiral module to the ANN.

Following \cite{field2014fast}, a dataset of $N=1000$ waveforms with mass ratios in the range $1\leq q \leq 2$ was generated and a surrogate model was built, with a tolerance of $10^{-10}$, resulting in a reduced basis of size $n=11$. Next, a simple symmetric encoder-decoder AE architecture was used, with a two-dimensional hidden representation and two hidden fully-connected layers of $128$ neurons on either side. The PReLU non-linearity \cite{he2015delving} was used in all layers. The model was built using the PyTorch Deep Learning framework \cite{pytorch}. The EIM coefficients were used as input and output for this network. The AE was trained for 100 epochs with an initial learning rate of $0.001$ and a batch size of 32. A multi-step multiplicative schedule was used with a gamma value of 0.9 and a step size of 15. The visual representation of the hidden layer is shown in Figure~\ref{fig:ae_hidden_rep}, with the colors indicating the $q$ values for each input coefficient. The spiral manifold in the hidden layer appears to describe a linear relationship between $q$ and the angle $\theta$ of the spiral. The mean squared error of the reconstruction is $6.82 \times 10^{-5}$.

\begin{figure}
    \centering
    \includegraphics[width=0.8\linewidth]{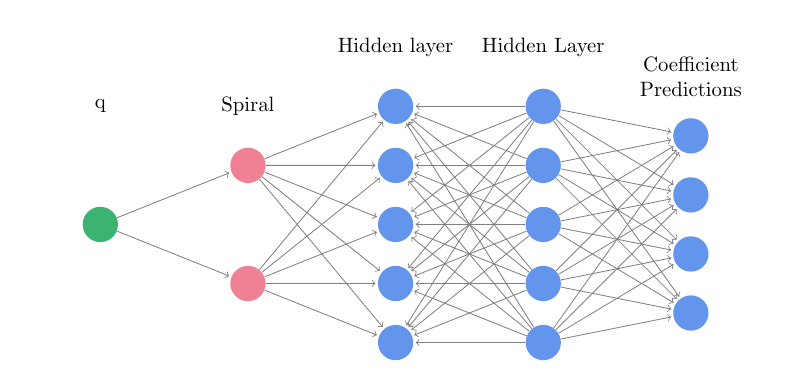}
    \caption{Fully connected neural network with two hidden layers and the included spiral module. Figure from \cite{2022Neurc.491...67N}.}
    \label{fig:nn_architecture}
\end{figure}

Based on the spiral pattern that emerged in Figure~\ref{fig:ae_hidden_rep}, a neural spiral module was proposed in \cite{2022Neurc.491...67N}, which first transforms the input $q$ into an angle $\theta$, defined as
\begin{equation}
    \label{eq:theta}
    \theta := w\cdot q + b,
\end{equation}
and subsequently maps $\theta$ onto a spiral structure of the form
\begin{align}
  \begin{split}
  \label{eq:spiral}
    s_x &:= (\alpha + \beta\cdot\theta) \cdot \cos\theta, \\
    s_y &:= (\alpha + \beta\cdot\theta) \cdot \sin \theta,
  \end{split}
\end{align}
where $w, b, \alpha$ and $\beta$ are parameters. These parameters are learnable, since the output is differentiable with respect to each of them. 
The spiral is fed to multiple, successive fully-connected layers, each with a nonlinear activation function, before reaching the final linear layer. An example of this architecture with two hidden layers is illustrated in Figure~\ref{fig:nn_architecture}. The inclusion of this module into an ANN accelerates the training process, leading to a significant reduction of the lowest achieved MSE.

The performance of various neural network architectures with fully-connected layers was assessed with and without the spiral module. The metrics used for evaluation were the waveform mismatch, inference speed, and memory requirements, with the maximum batch size that can be processed in a single forward pass on an NVIDIA RTX 2080 Ti GPU. All networks were trained for 2500 epochs with a batch size of 16, using the Adam optimizer \cite{kingma2014adam} and an initial learning rate of 0.001, which was reduced by 0.95 every 150 epochs.

The inclusion of the spiral module significantly improved the mismatch achieved. When only one hidden layer was used, the baseline network with 128 neurons produced waveforms with a very poor mismatch ($1.03 \times 10^{-1}$ median mismatch). However, with the addition of the spiral module, even with only 32 hidden neurons, the median mismatch decreased by about 6 orders of magnitude. The best median and $95^{\rm th}$ percentile mismatch ($9.41 \times 10^{-9}$ and $3.48 \times 10^{-8}$) was achieved by the $\mathcal{S}$-32-64-128-64 network, which was able to generate up to 3.4 million coefficients in a single forward pass on the aforementioned GPU.

\begin{figure}[t]
    \centering
    \includegraphics[width=0.9\textwidth]{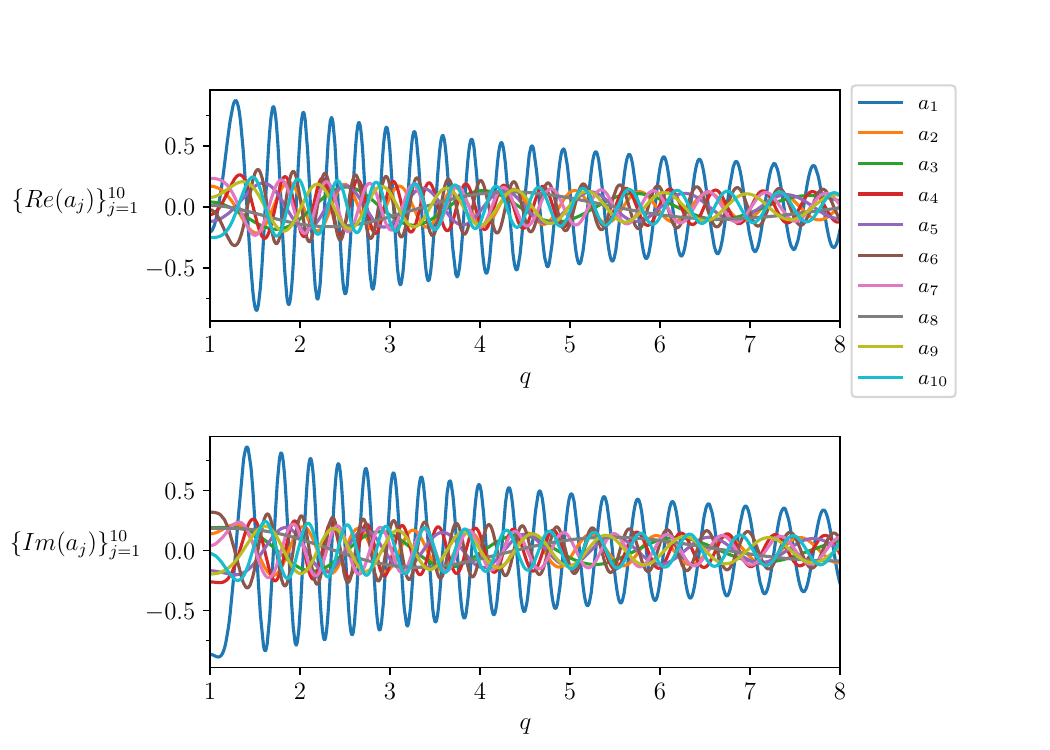}
    \caption{Real (top) and imaginary parts (bottom) of the empirical interpolation coefficients $a_j(q)$ for a surrogate model of EOBNRv2 waveforms that is valid for $1\leq q \leq 8$. Figure from \cite{2022Neurc.491...67N}.}
    \label{fig:coeffs_q1to8}
\end{figure}

Finally, a spiral module was added to a neural network that was trained on a larger dataset of $N=56000$ waveforms with $1\leq q \leq 8$, where the $q$ values were equidistant. A validation and a test set were also created, each with $14000$ waveforms, and the $q$ values were randomly chosen in the range of $1\leq q \leq 8$. Figure~\ref{fig:coeffs_q1to8} shows the real and imaginary parts of the first ten coefficients of the EIM basis, $\{a_j(q)\}_{j=1}^{10}$. Despite some modulation of amplitude, each coefficient has a sinusoidal dependence with $q$ (except near $q=1$, where $dq/da_j=0$ for all $j$).

\begin{figure}[t]
    \centering
    \includegraphics[width=0.9\linewidth]{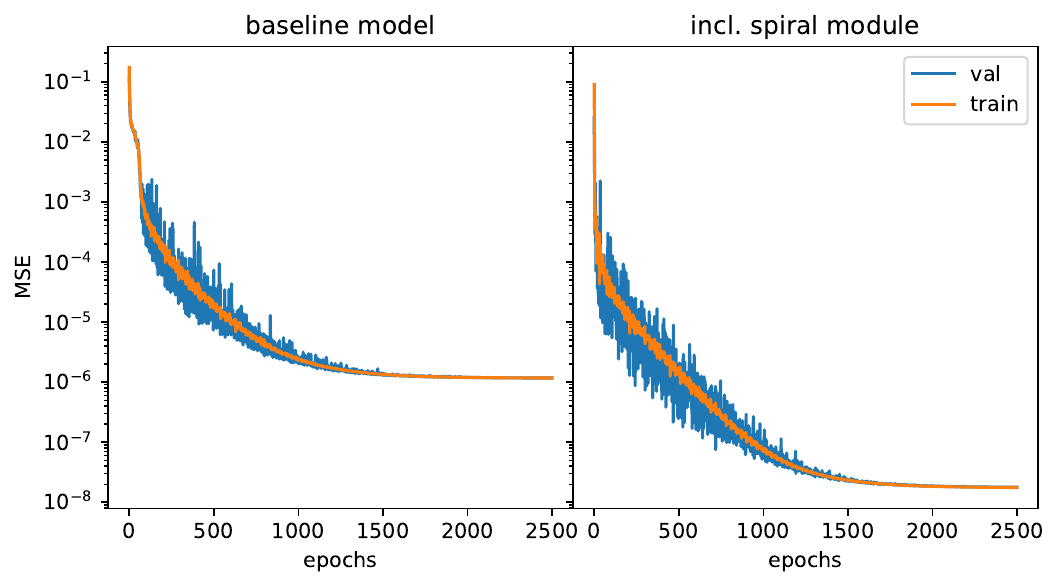}
    \caption{Training and validation loss per epoch for the 32-64-128-64 network and the corresponding architecture with the addition of the spiral. Figure from \cite{2022Neurc.491...67N}.}
    \label{fig:spiralloss}
\end{figure}

Several neural networks were trained and tested on the dataset. All networks were trained for 5000 epochs, with a batch size of 32, and the Adam optimizer \cite{kingma2014adam} with an initial learning rate of 0.001, which was reduced by 0.9 every 30 epochs. The training and validation loss per epoch for the $32-64-128-64$ network and the corresponding architecture with the addition of the spiral is shown in Figure~\ref{fig:spiralloss}. The spiral addition resulted in a lower mean squared error, allowing smaller networks to achieve the same accuracy as larger networks, leading to a larger batch size that can be processed in a single forward pass when using a specific GPU card.

\section{GW Detection with Deep Residual Networks}

The fourth observing run (O4) of gravitational wave detectors, which began in spring of 2023, is expected to result in a significant increase in the number of detections. This increase will be even more pronounced during O5 and the observing runs of the planned third-generation detectors (e.g. Cosmic Explorer \cite{2019BAAS...51g..35R} and Einstein Telescope \cite{2020JCAP...03..050M}). However, the application of traditional matched-filtering techniques to obtain near real-time detection triggers is becoming increasingly costly or even impractical \cite{2021arXiv211106987C}, due to both computational efficiency and accuracy. This is especially true for near-threshold systems with random spin directions, which require a much larger parameter space than the aligned-spin case. The situation will become even more difficult if template banks with departures from general relativity (GR) are included. Unmodeled search algorithms, on the other hand, have limited sensitivity, depending on the particular GW source.

Recently, the implementation of machine-learning (ML) methods, such as convolutional neural networks (CNN) or auto-encoders, has been investigated as an attractive solution to the problem of detecting gravitational waves (GWs), see e.g. 
\cite{PhysRevLett.120.141103,PhysRevD.97.044039,2019PhRvD.100f3015G,CORIZZO2020113378,PhysRevD.102.063015,2020PhRvD.101j4003W,2020PhLB..80335330K,2020arXiv200914611S,2021PhRvD.103f3034L,2021NatAs...5.1062H,2021MNRAS.500.5408M,2021PhLB..81236029W,2021PhRvD.104f4051J,10.3389/frai.2022.828672,2022arXiv220208671C,PhysRevD.105.043003,2022arXiv220606004B,schafer2022training,PhysRevD.106.042002,2022arXiv220704749A,2022arXiv220612673V,2022PhRvD.106h4059A, 2022MNRAS.516.3847G, 2023PhRvD.107h2003A,2023PhRvL.130q1402L,2023PhRvL.130q1403D,2023CQGra..40m5008B,2023arXiv230615728T,2023PhRvD.108d3024M,2023MLS&T...4c5024B,2022arXiv220111126M,2023PhLB..84037850Q,2023arXiv230519003J,2023CQGra..40s5018F,2023arXiv230716668G} and \cite{cuoco2020review,app13179886,2023arXiv231115585Z} for reviews. However, it has been difficult to evaluate the effectiveness of such efforts in a realistic setting. The first Machine-Learning Gravitational-Wave Mock Data Challenge (MLGWSC-1) was completed \cite{challenge1}, providing an objective framework for testing the sensitivity and efficiency of ML algorithms on modeled injections in both Gaussian and real O3a detector noise in comparison to traditional algorithms. In \cite{2023PhRvD.108b4022N},  the leading ML algorithm in the case of real O3a noise was presented in more detail and it was shown that with further improvements it surpasses, for the first time, the results obtained with standard configurations of traditional algorithms in this specific setting. This was achieved for a component mass range between $7-50 M_\odot$ (which corresponds to $70\%$ of the announced events in the cumulative GWTC catalog \cite{GWTC3}) and a relatively low false-alarm rate (FAR) as small as one per month.

\begin{figure*}[t!]
    \centering
    \includegraphics[width=\textwidth]{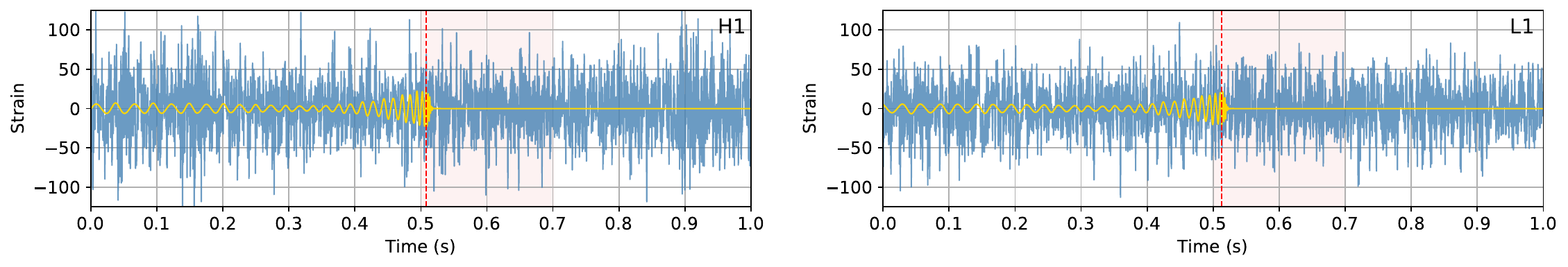}
    \caption{A 2-channel segment of data from the training set used in \cite{2023PhRvD.108b4022N} is displayed in the left and right panels, respectively, for the Hanford (H1) and Livingston (L1) detectors. The whitened strain of a 1 s segment is shown around the time of coalescence. The coalescence times in the detector frames are within the 0.5-0.7 second range (indicated by the shaded area). The injected waveform is scaled to match the difference between the whitened foreground and background segments. Figure from \cite{2023PhRvD.108b4022N}.}
    \label{fig:injections}
\end{figure*}

The AresGW algorithm, described in \cite{2023PhRvD.108b4022N}, combines several components that increase the sensitive distance. It is based on a 54-layer one-dimensional deep residual network (ResNet) \cite{he2016deep}, which is more capable than a simpler CNN. Additionally, the Deep Adaptive Input Normalization (DAIN) \cite{dain} was included to address the non-stationary nature of real O3a noise. Furthermore, the dataset was augmented during training to improve the results. The execution speed was increased with the implementation of a framework-specific, module-based whitening layer, which computes the power spectral density (PSD) in batched tensor format. Finally, curriculum learning was used, which allowed the network to learn waveforms with the highest signal-to-noise ratio (SNR) first.
The network was created using PyTorch \cite{PyTorch_ref} and trained (including validation) on 12 days of data in 31 hours on an A6000 GPU (for 14 epochs). The evaluation of one month of test data on the same hardware took less than 2 hours. The main findings of \cite{2023PhRvD.108b4022N} are summarized below.

\subsection{Training and test datasets}
\label{sec:datasets}

The training dataset in \cite{2023PhRvD.108b4022N} spanned a period of 12 days and included real noise from the O3a LIGO run and injections of non-aligned binary black hole waveforms (in accordance with dataset 4 in \cite{challenge1}). Noise was taken from sections of O3a that are available from the Gravitational Wave Open Science Center (GWOSC) \cite{GWOSC}. Only segments with a minimum length of 2 hours, where both LIGO detectors had good quality data, and excluding 10 seconds around detections listed in GWTC-2 were included (see \cite{challenge1} for more information). Applying these criteria, the dataset had noise from each of the two aLIGO detectors, Hanford (H1) and Livingston (L1), with a total duration of 11 weeks and a sampling rate of 2048 Hz.

The waveforms injected into the training set were generated using the IMRPhenomXPHM waveform model \cite{IMRPhenomXPHM_ref}, with a lower-frequency cutoff of 20 Hz. The masses of the individual components, $m_1$ and $m_2$, ranged from 7 to 50 solar masses, resulting in a maximum signal duration of 20 seconds. The signals were uniformly distributed in coalescence phase, polarization, inclination, declination, and right ascension (see \cite{challenge1} for more information). The waveforms were not uniformly distributed in volume, but instead, the chirp distance $d_c$ \cite{challenge1} was sampled (as opposed to the luminosity distance $d$). This selection increases the number of low-mass systems that can be detected. The spins of the individual components had an isotropically distributed orientation with a magnitude between 0 and 0.99, which means that precession effects were present. All higher-order modes up to $(4,-4)$ available in IMRPhenomXPHM were included. A representative data segment of the training set is shown in Figure \ref{fig:injections}.

The training data set was made up of the first 12 days of the 11-week dataset, resulting in 740k noise segments that formed the background noise. A set of 38k different waveforms was randomly injected into around 19 different background segments each, resulting in 740k foreground segments that contain injections, creating a balanced training set. Additionally, a validation set was created, based on weeks 4 to 7 of the 11-week dataset and with a different random seed for injections. Lastly, the test data set consisted of noise from weeks 8-11 of the 11-week dataset and injections with merger times randomly spaced between 24s to 30s. For the test dataset the same random seed and offset as in \cite{challenge1} was used.


\begin{figure*}[t!]
    \centering
    \includegraphics[width=\textwidth]{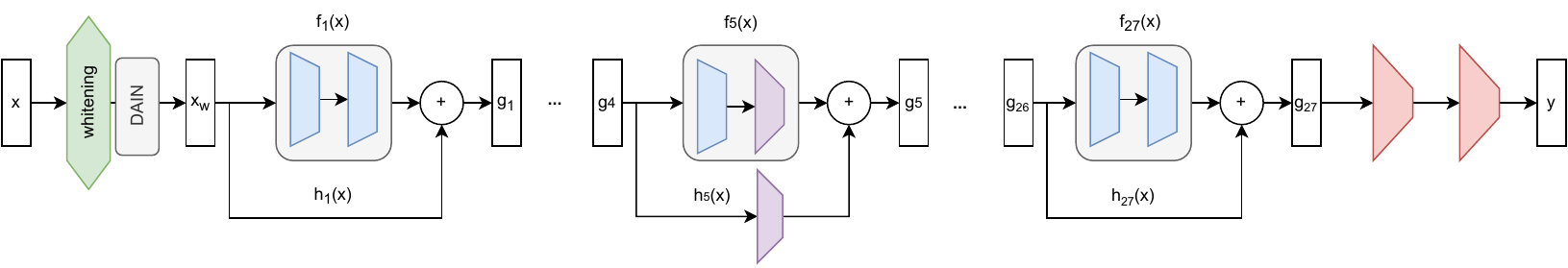}
    \caption{Description of the residual network architecture in \cite{2023PhRvD.108b4022N}. The input $\mathbf{x}$ is $2\times 2048$-dimensional (see the text for details). Figure from \cite{2023PhRvD.108b4022N}.}
    \label{fig:resnet52}
\end{figure*}

The training data was first pre-processed using whitening as in \cite{usman2016pycbc,PhysRevLett.120.141103} and then normalized using the DAIN algorithm \cite{dain,passalis2021forecasting}. DAIN  is trained by back-propagating the network’s gradients to its parameter. Furthermore, DAIN can adjust the normalization scheme applied to the input \emph{during inference}, thus allowing it to handle non-stationary data.

\subsection{Deep Residual Networks}

Residual neural networks \cite{he2016deep} employ skip connections to improve training, allowing gradients to better reach the earliest layers of the neural network architecture, thus solving the problem of vanishing gradients \cite{hanin2018neural}. This, in combination with well-crafted training methods \cite{wightman2021resnet}, leads to more effective training as the number of layers increases. This allows for the training of much deeper networks than simple CNNs.

The deep residual network developed in \cite{2023PhRvD.108b4022N} was based on 1D convolutions for the binary classification of 1s long (i.e. $2\times 2048$-dimensional) segments into either positive (containing an injection) or negative (pure noise) segments. The network had a depth of 54 layers, which were grouped into 27 blocks containing two convolutional layers with a varying number of filters. Blocks 5, 8, 11, 14, and 17 were 2-strided, which means that the dimensionality was halved and an additional layer was used in the residual connection. Each convolutional layer was followed by batch normalization and a ReLU activation function. The two final individual convolutional layers reduced the output to a binary outcome (noise plus injected waveform vs. noise only). The Adam optimizer \cite{https://doi.org/10.48550/arxiv.1412.6980} was used for backpropagation and regularized binary cross entropy (a variant of the finite cross-entropy loss function \cite{schafer2022training}) was used as the objective function. During training, dynamic augmentation was also employed. A graphical representation of the network architecture is shown in Figure~\ref{fig:resnet52}.

\begin{figure}
    \centering
    \includegraphics[width=0.7\textwidth]{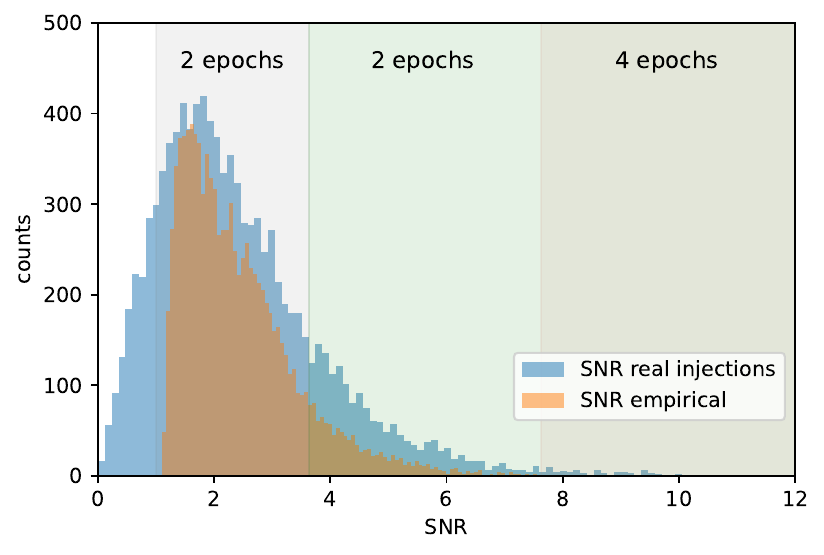}
    \caption{Comparison of the SNR histogram, calculated with the empirical relation in Eq.~(\ref{SNR-empirical}), to the optimal SNR of $10^4$ randomly chosen injections. The shaded areas in the graph represent the limited SNR values used in the first eight epochs of the {\it learning strategy}. Figure from \cite{2023PhRvD.108b4022N}}
    \label{fig:SNR_comparison}
\end{figure}

 A learning strategy was developed such that the network was initially trained on the strongest injections and only later on weaker injections. This is accomplished by using the optimal signal-to-noise ratio (SNR) of the injected signal, which is given by the equation \begin{equation} {\rm SNR} =2 \sqrt{ \int_0^{\infty} df \frac{ \tilde{h}(f)^2}{S_n(f)}}, \label{SNR-optimal} \end{equation} 
where $\tilde{h}(f)$ is the amplitude of the Fourier transform of the injected signal and $S_n(f)$ is the power spectral density of the detector noise. Instead of using the actual optimal SNR, an empirical relation was created that only depends on the chirp mass, the distance, and the inclination angle  $\iota$
\begin{equation}
\mathrm{SNR}=\frac{1261 \ {\rm Mpc}}{D} \left(\frac{{\cal M}_{\rm c}}{M_\odot}\right)^{5 / 7}\left[0.7+0.3\cos (2 \ \iota)\right].
\label{SNR-empirical}
\end{equation}
Figure \ref{fig:SNR_comparison} displays a comparison between the optimal SNR (calculated using Eq. (\ref{SNR-optimal}) and the PSD of the Hanford detector) for a sample of $10^4$ randomly chosen injections and the approximate SNR as determined by Eq. (\ref{SNR-empirical}). The two distributions are similar; however, the actual SNR is affected by a variety of other factors (such as sky location and spins).

The training process began with signals that were easily recognizable and had a high estimated SNR for the first four epochs. Subsequently, the network was gradually trained on weaker signals, and after the tenth epoch, it learned all the signals in the training set. Figure \ref{fig:SNR_comparison} shows the first eight epochs compared to the SNR distribution. As a result of the chosen learning strategy, initial losses were low. When the network had learned all the signals (after ten epochs), the loss of the training set was equal to that of the validation set.

\subsection{Detection of BBH injections in real noise}

\begin{figure}[t]
	\centering
	\includegraphics[width=0.7\textwidth]{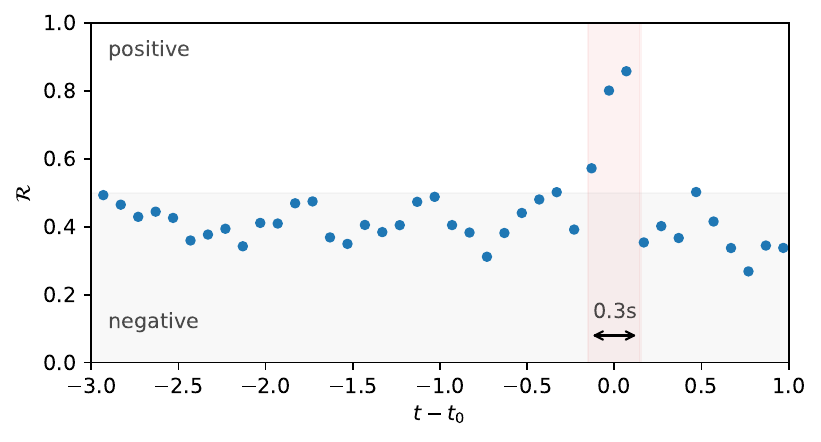}
	\caption{Representative example of the ranking statistic for overlapping segments (with a stride of 0.1s). Segments with ${\cal R}<0.5$ are classified as negatives. Segments with ${\cal R}>0.5$ that cluster within 0.3s of a known injection at time $t_0$ are reported as a single true positive. Figure from \cite{2023PhRvD.108b4022N}}
	\label{fig:trigger}
\end{figure}

The trained network was used to analyze segments of the test dataset, producing a binary output that corresponds to the probability that an injection is present or that the segment only contains noise. The first output was used as a ranking statistic $\cal R$ (with values between 0 and 1). When ${\cal R}>0.5$, a positive result was recorded. Clustering was applied to any positives that were detected within a time interval of 0.3 s, which were reported as a single detection (see Fig. \ref{fig:trigger} for an example). After deployment, the output was evaluated every 0.1 seconds and compared with the known injection times in the test dataset. If a positive output was within 0.3 seconds of the nominal merger time for a particular injection, it was classified as a true positive and, otherwise, as a false positive.

In order to assess the effectiveness of the search algorithm, one first calculates the false alarm rate as a function of the ranking statistic, that is, the ${\rm FAR} (\cal R)$ function. Subsequently, the sensitivity of the search is determined as a function of the ranking statistic, which produces a relationship between the sensitivity and FAR (see \cite{challenge1} for definitions and the complete methodology).

\begin{figure}[t]
    \centering
    \includegraphics[width=0.8 \linewidth]{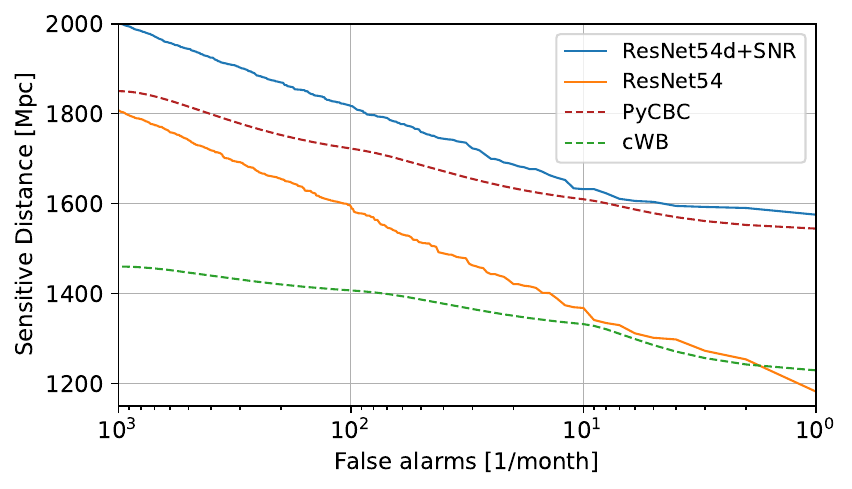}
    \caption{Comparison of the  performance of the best model (ResNet54d+SNR) in \cite{2023PhRvD.108b4022N} with the simpler setup (ResNet54) used in \cite{challenge1} and two leading algorithms for GW detection, Coherent WaveBurst (cWB) and PyCBC (which only used aligned-spin templates, see text). All codes were tested on the same dataset established in \cite{challenge1}. The ResNet54d+SNR model outperformed the other algorithms at all false alarm rates in this setting. Figure from \cite{2023PhRvD.108b4022N}.}
    \label{fig:comparison}
\end{figure}

Fig. \ref{fig:comparison} shows the sensitivity distance as a function of FAR for the best model (ResNet54d+SNR), compared to a simpler setup (ResNet54) and two leading algorithms for GW detection, Coherent WaveBurst (cWB) and PyCBC. cWB is a waveform model-agnostic search pipeline for GW signals based on the constrained likelihood method \cite{PhysRevD.93.042004,cwb-softwareX,klimenko_sergey_2021_5798976}, while PyCBC \cite{alex_nitz_2022_6912865} is based on a standard configuration of archival search for compact-binary mergers \cite{2021arXiv211206878N}. The cWB and PyCBC results presented in Figure \ref{fig:comparison} are taken from \cite{challenge1}, where they were obtained on the same test dataset. cWB uses wavelets, which prevents it from achieving ideal fitting factors. It was recently improved with machine-learning techniques \cite{PhysRevD.105.083018}. PyCBC implements matched filtering of waveform templates, but in \cite{2021arXiv211206878N,challenge1} only aligned-spin templates were used (for more general waveforms, the method could become computationally too expensive). Since the injections in the test dataset of \cite{2023PhRvD.108b4022N} were based on more general waveforms, this particular PyCBC search could not reach ideal fitting factors, leaving room for other algorithms to outperform it. As seen in Figure \ref{fig:comparison}, the best model presented in \cite{2023PhRvD.108b4022N}, which included SNR-based curriculum learning, outperformed the PyCBC results at all FAR and also significantly surpassed the sensitivity of the unmodeled cWB search.

\section{ANN Surrogate Models of Neutron Star Mass-Radius Relations in Alternative Theories of Gravity}
    \label{sec:ANNalt}
    
 Several studies have used Artificial Neural Networks (ANNs) to reconstruct the EoS of neutron stars (NSs) based on their observable properties \cite{PhysRevD.101.054016,PhysRevD.98.023019,Fujimoto_2021,Ferreira_2021,galaxies10010016}. For instance, \cite{morawski_2020} investigated the use of ANNs with the autoencoder architecture, while \cite{Soma_2022,Soma_2023} employed ANNs to represent the EoS in a model-independent way, utilizing the unsupervised automatic differentiation framework. Other machine-learning techniques have been applied to explore the NS EoS. For example, \cite{Lobato_2022} used a clustering method to identify patterns in mass-radius curves, and \cite{lobato2022unsupervised} investigated correlations among different EoSs of dense matter using unsupervised Machine Learning (ML) techniques. Additionally, attempts have been made to derive nuclear matter characteristics from NS EoS and observations using deep neural networks; see, e.g., \cite{Ferreira_2022,krastev2023deep}.
 
   Neutron stars have been the focus of theoretical investigations in alternative theories of gravity (see \cite{2022arXiv221101766D,Berti_2015} for reviews and possible tests, and \cite{Charmousis_2022} for the particular theory discussed in this work). Bayesian statistics is commonly used to infer the NS EoS from observations of their macroscopic properties. This requires a TOV (Tolman-Oppenheimer-Volkoff) solver to be run multiple times to obtain the final posterior distribution of various parameters. If modified theories of gravity are to be included in these studies, a modified TOV solver, such as the one presented in \cite{Charmousis_2022}, is needed. However, this algorithm is based on an iterative method for solving a differential equation system, which makes Bayesian inference computationally expensive. Therefore, it would be beneficial to find an alternative way to quickly and accurately predict the macroscopic properties of NSs, given some defining characteristics or other macroscopic properties of each equilibrium model.

\begin{figure*}[t!]
        \centering
        \includegraphics[width=1.0\textwidth]{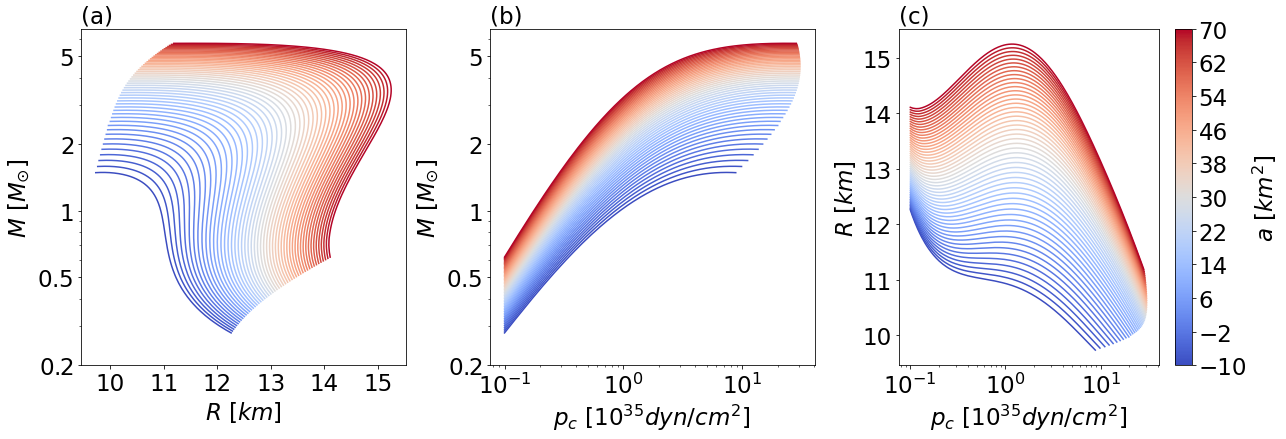}
        \caption{Data set for EoS BSk20, for modeling the function $f_1$. Figure from \cite{2023arXiv230903991L}.}
        \label{fig:alldata}
    \end{figure*}
    
Motivated by the need for a faster process, we developed an Artificial Neural Network (ANN) regression for two types of functions, $f_1({\rm{EoS}}; \alpha, p_c) \rightarrow (M,R)$ and $f_2({\rm{EoS}}; \alpha, M) \rightarrow R$, which was implemented in \cite{2023arXiv230903991L} and used in a Bayesian inference application in \cite{2023arXiv230905420B}. The EoS is a distinct variable, with each type having one ANN model for each EoS. $\alpha$ is the coupling constant of the theory and $p_c$ is the central pressure of the NS. The first type serves as a surrogate model for the numerical iterative method described in \cite{Charmousis_2022}, which provides the mass and radius of NSs for a specific EoS and a given pair of $\alpha$ and $p_c$. The aim is to speed up the process while still meeting strict accuracy requirements. The second type cannot be obtained directly using the iterative method since $p_c$ must be an input. Therefore, implementing a root-finding algorithm would be the only solution, resulting in further time delays. On the other hand, training ANNs to predict $R$ based on $(\rm{EoS}; \alpha, M)$ offers a more straightforward approach to handling type $f_2$.

 The 4D Horndeski scalar-tensor model, which was studied in \cite{2023arXiv230903991L}, was derived from higher-dimensional Einstein-Gauss-Bonnet gravity. The action of this model included several nonlinear terms of a scalar field $\phi$, and the mass and radius of the neutron star were affected by the strength of the coupling constant $\alpha$, which has units of length square. If $\alpha=0$, then the model is equivalent to Einstein's General Relativity. Further information can be found in \cite{Charmousis_2022}.
    
     \begin{table}[]
        \centering
            \caption{Final network architecture in  \cite{2023arXiv230903991L}.}
        \label{tab:architecture}
        \begin{tabular}{cccr}
        \toprule
            Layer \hspace{0.2cm} & \hspace{0.2cm} Type $f_1$ \hspace{0.2cm} & \hspace{0.2cm} Type $f_2$ \hspace{0.2cm}\\
            \midrule
            Input layer     &  $(\alpha,p_c)$   & $(\alpha,M)$\\
            Hidden layer 1  &  25-tanh          & 25-tanh\\
            Hidden layer 2  &  35-relu          & 35-relu\\
            Hidden layer 3  &  25-tanh          & 25-tanh\\
            Output layer    &  $(M,R)$          & $R$ \\
            \bottomrule
        \end{tabular}
    \end{table}
    
    \begin{figure*}[t!]
        \centering
        \includegraphics[width=1.0\textwidth]{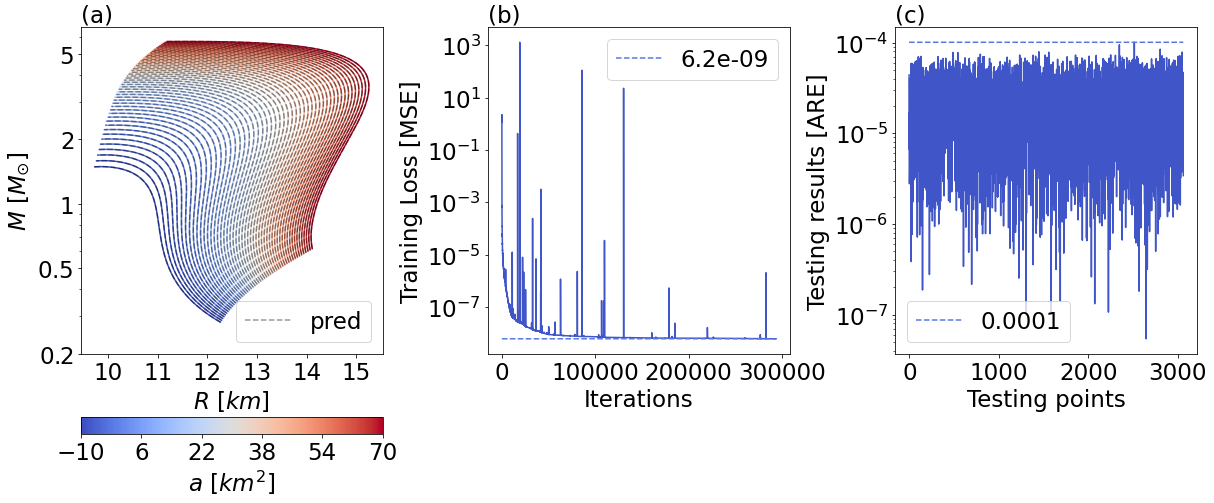}
        \caption{Training results for $f_1$ and EoS BSk20. Figure from \cite{2023arXiv230903991L}.}
        \label{fig:eos_bsk20}
    \end{figure*}

\section{Training and Testing}
    \label{sec:traintest}
   The numerical code from \cite{Charmousis_2022} requires two inputs to generate the data sets in \cite{2023arXiv230903991L}: the coupling constant $\alpha \: [\rm{km}^2]$ of the theory and the central pressure $p_c \: [10^{35} \: \rm{dyn}/\rm{cm}^2]$. These inputs are used to calculate the mass $M \: [\rm{M}_\odot]$ and radius $R \: [\rm{km}]$ of the neutron star. For each type of function, 20 data sets were created, each based on one of the 20 tabulated equations of state.
    
    \begin{figure*}[t!]
        \centering
        \includegraphics[width=0.47\linewidth]{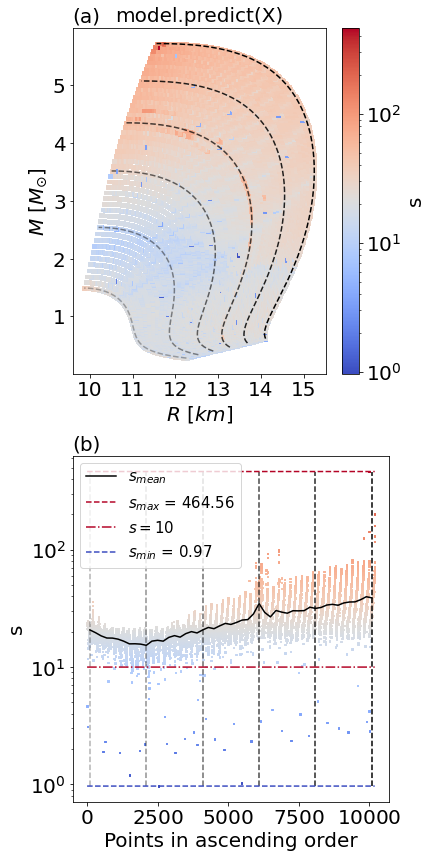}
        \includegraphics[width=0.47\linewidth]{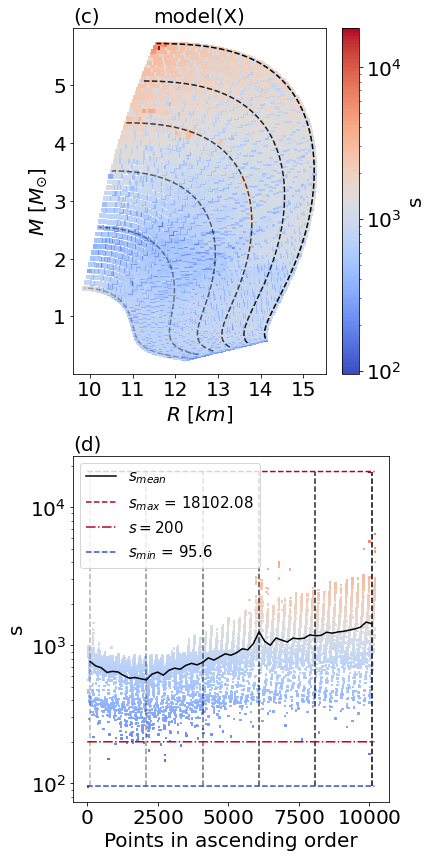}
        \caption{Speed-up per $M-R$ data point showing how many times faster the trained ANN model is than the numerical code in each case. Figure from \cite{2023arXiv230903991L}.}
        \label{fig:f1_ratio}
    \end{figure*}

    The type of function $f_1(\rm{EoS}; \alpha, p_c)$ maps a given set of 51 values of $\alpha$ and 200 values of $p_c$ to a unique pair of $(M,R)$ values. The $\alpha$ values were evenly distributed on a linear scale, while the $p_c$ values were logarithmically spaced between 0.1 and 1.2 times the central pressure required for a NS to reach its maximum mass for a specific value of the coupling constant. As an example, Figure \ref{fig:alldata} shows the mapping of $(\alpha, p_c)$ pairs to $(M,R)$ pairs for EoS BSk20. The type of function $f_2(\rm{EoS}; \alpha, M) \rightarrow R$ had the same size as $f_1$, but the range of $p_c$ values was chosen differently. The data comprised 51 values of $\alpha$ ranging from -10 to 70, and 200 values of $p_c$ ranging from 0.1 to $p_{\rm{max}}$.

   The TensorFlow module Keras\footnote{\texttt{\url{https://www.tensorflow.org/api_docs/python/tf/keras}}} was employed with a 70:30 train-test ratio. The Mean Square Error (MSE) was selected as the loss function, while the Absolute Relative Error (ARE) was used as the criterion for evaluating the trained models. The ANN architecture for each type is presented in Table \ref{tab:architecture}. The second-order {\it Broyden–Fletcher–Goldfarb–Shanno optimizer} algorithm (BFGS), as proposed in \cite{e25010175}, was implemented as the preferred optimizer\footnote{BFGS is not included in the list of provided Keras optimizers.}.
    
   Fig. \ref{fig:eos_bsk20}a provides an indication of the training results for $f_1$, comparing the actual output with the predicted output. Fig. \ref{fig:eos_bsk20}b displays the training loss per iteration in terms of the Mean Square Error (MSE) and Fig. \ref{fig:eos_bsk20}c presents the Absolute Relative Error (ARE) at a given test point $i$ 
\begin{equation}
\mathrm{ARE}_i=\frac{1}{m} \sum_{j=1}^m\left|\frac{Y_i^j-Y_{i, \text { true }}^j}{Y_{i, \text { true }}^j}\right|
\end{equation}
   where $m$ is the number of output neurons, $Y_i^j$ is the {\color{black}network output with index $j$ for test point with index $i$}, and $Y_{i,\rm{true}}^j$ is the corresponding real output. The Mean ARE over all $n=3060$ points in the test dataset is denoted as MARE. The MARE ranged between $10^{-5}$ and $10^{-4}$ for all EOSs, while the maximum ARE was $6\times 10^{-4}$ for $f_1$ and $9\times 10^{-3}$ for $f_2$, indicating that the absolute relative error never exceeded 1\% in the entire domain of the training and test sets.
   
    \subsection{Speed-up}
       The \textit{speed-up} $s$ when comparing the speed of the trained ANN models of $f_1$ to the speed of the numerical code, is defined as:
        \begin{equation}
            s = \frac{\Delta t_{\rm{ANN}}}{\Delta t_{\rm{num}}},
        \end{equation}
        {\color{black}where $\Delta t_{ANN}$ is the run time when using an ANN model and $\Delta t_{\rm{num}}$ the run time of the numerical code with the iterative numerical scheme.}  The output of the models can be calculated in three different ways, leading to a different  speed-up:
        \begin{enumerate}
            \item \texttt{model.predict($X$)}, with $X$ being one input value,
            \item \texttt{model($X$)}, with $X$ being one input value,
            \item \texttt{model.predict($\bf X$)}, with $\bf X$ being an array of input values.
        \end{enumerate}
        
           The left panel of Figure \ref{fig:f1_ratio} shows the speed up for \texttt{model.predict($X$)}. Panel \ref{fig:f1_ratio}a displays the speed up using different colors, and the dashed lines with increasing transparency represent six different $M-R$ curves for $a = \{-10,6,22,38,54,70\}$ $\rm km^2$ respectively.
Panel \ref{fig:f1_ratio}b shows the speed-up values for a particular arrangement of the data points, which are sorted in ascending order of $\alpha$ and $p_c$. Most of the speed-up values were between 10 and 100, with less than 1\% outside of this range.
            Generally, larger input values tend to result in higher speed-ups, with an average speed-up of approximately 25 across all data points.
            
          The right panel of Figure \ref{fig:f1_ratio} shows the acceleration for the \texttt{model(X)} case, which is set up similarly to the left panel.  Most of the speed-up values are between 200 and 18000 (less than 0.2\% are outside of this range). The average speed-up, taking into account all data points, was greater than 900. For the \texttt{model.predict($\bf X$)} case, the 10200 data points were entered as an array of input values $\bf X$. In this case,  the \textit{effective run time} $\Delta t_{eff}$, {\color{black} can be calculated, which is defined as 
            \begin{equation}
                \Delta t_{eff} = \frac{\Delta t_{N}}{N},
            \end{equation}
where $\Delta t_{N}$  is the {\it total run time} for the whole array as input and $N = 10200$. The speed-up ranged from $\sim 4600$ to $\sim 565000$, with a mean value of $\sim 31000$. 

\section{Conclusions}

This comprehensive review has highlighted several advancements in the field of gravitational wave astronomy through the application of machine learning techniques. The exploration of various neural network architectures, including deep residual networks and autoencoders, has demonstrated the potential for improvements in the detection and analysis of gravitational wave signals. The successful integration of machine learning in this domain not only concerns the optimization of data analysis, but also can open new avenues for understanding complex astrophysical phenomena. Moreover, the application of machine learning in neutron star mass-radius relation studies in alternative theories of gravity could be used to break the degeneracy between the equation of state and some alternative theories of gravity. 

In conclusion, the integration of machine learning techniques in gravitational wave astronomy and neutron star studies represents a paradigm shift. Not only does it enhance the accuracy and efficiency of existing methodologies, but it also paves the way for novel discoveries in the realm of compact object astrophysics. The continued collaboration between the gravitational wave and machine learning communities is vital for further advancements.

\begin{acknowledgement}

I am grateful to my collaborators, Theocharis Apostolatos, Stella Fragkouli, Panagiotis Iosif, Alexandra Koloniari, Ioannis Liodis, Paraskevi Nousi, George Pappas, Nikolaos Passalis,  Evangelos Smyrniotis, and Anastasios Tefas, for their contributions that led to the main publications summarized in this review. Many thanks to Elena Cuoco for comments on the manuscript. This work was carried out within the framework of the EU COST action No. CA17137.
This research has made use of data or software obtained from the Gravitational Wave Open Science Center (gwosc.org), a service of the LIGO Scientific Collaboration, the Virgo Collaboration, and KAGRA. This material is based upon work supported by NSF's LIGO Laboratory which is a major facility fully funded by the National Science Foundation, as well as the Science and Technology Facilities Council (STFC) of the United Kingdom, the Max-Planck-Society (MPS), and the State of Niedersachsen/Germany for support of the construction of Advanced LIGO and construction and operation of the GEO600 detector. Additional support for Advanced LIGO was provided by the Australian Research Council. Virgo is funded, through the European Gravitational Observatory (EGO), by the French Centre National de Recherche Scientifique (CNRS), the Italian Istituto Nazionale di Fisica Nucleare (INFN) and the Dutch Nikhef, with contributions by institutions from Belgium, Germany, Greece, Hungary, Ireland, Japan, Monaco, Poland, Portugal, Spain. KAGRA is supported by Ministry of Education, Culture, Sports, Science and Technology (MEXT), Japan Society for the Promotion of Science (JSPS) in Japan; National Research Foundation (NRF) and Ministry of Science and ICT (MSIT) in Korea; Academia Sinica (AS) and National Science and Technology Council (NSTC) in Taiwan. 

\end{acknowledgement}

\bibliographystyle{unsrt}

\end{document}